\documentclass[oneside,a4paper,english,links,12pt]{article}

\pdfoutput=1
\usepackage{amsmath}
\usepackage{authblk}
\usepackage{natbib}

\usepackage{amssymb}
\usepackage{subfigure}
\usepackage{graphicx}
\usepackage{amsfonts}

\usepackage{calc}
\usepackage{ifpdf}
\usepackage{indentfirst}
\usepackage{authblk}
\usepackage{babel}
\usepackage{color}
\usepackage{hyperref}
\usepackage{nameref}
\usepackage{url}
\usepackage{times}
\usepackage{fontenc}
\usepackage{hyperref}

\usepackage{vmargin}

\setpapersize{A4}
\setmargins{2.5cm}       % margen izquierdo
{1.5cm}                        % margen superior
{16.5cm}                      % anchura del texto
{23.42cm}                    % altura del texto
{12pt}                           % altura de los encabezados
{1cm}                           % espacio entre el texto y los encabezados
{0pt}                             % altura del pie de página
{2cm}                           % espacio entre el texto y el pie de página

\def\brokenline{{$- - - $}} 
 
\def\lines{{$- . - . -$\small}}
\def\bullets{{\scriptsize$\bullet \cdot \bullet \cdot \bullet$\small}}
\def\diamonds{\normalsize{$\diamond\cdot\diamond\cdot\diamond$}\small}
\def\circs{\normalsize{$\circ \cdot \circ \cdot \circ\;$\small}}

\def\boxs{\scriptsize{$\Box \cdot \Box \cdot \Box $ \small}}

\def\pluss{\scriptsize{$+\cdot+\cdot+$}\small} 
\title{Modeling Turbulent Flows with LSTM Neural Network}

\author[1]{Hugo D. Pasinato}

\author[2]{Nicol\'as F. Moguilner Rhe}

\affil[1,2]{(Universidad Tec. Nacional, FRP, Argentina)}

\affil[1]{hugopasinato@frp.utn.edu.ar}

\date{}

\begin{document}

\maketitle 

\noindent In this study, we explore the application of an artificial
recurrent neural network (RNN) called Long Short-Term Memory (LSTM) as
an alternative to a turbulent Reynolds-Averaged Navier-Stokes (RANS)
model. The LSTM models are utilized to predict the shear Reynolds
stress in developed and developing turbulent channel flows. We conduct
comparative analyses, comparing the LSTM results propagated through
computational fluid dynamics (CFD) simulations with the outcomes from
the $\kappa-\epsilon$ model and data acquired from direct numerical
simulation (DNS). These analyses demonstrate a good performance of the
LSTM approach.

\section{Introduction}

Fluid flow predictions are of utmost importance in various fields of engineering, including mechanical, chemical, and aeronautical, to achieve diverse objectives such as design optimization and equipment enhancement. Many of these engineering applications involve turbulent flows, which require the
use of Reynolds-averaged Navier-Stokes (RANS) modeling. RANS is a
commonly employed technique for high-velocity scenarios with high
Reynolds numbers. It involves utilizing the Reynolds averaging method
to derive the time-averaged general equations of fluid motion, known
as the RANS equations (\cite{pope2000}). These equations, expressed in dimensionless form using tensor notation, are the conservation of mass

\begin{equation}\label{eq101}
\frac{\partial U_i}{\partial x_i } =0,
\end{equation}

\noindent and the momentum equations

\begin{equation}\label{eq102}
\frac{\partial U_i}{\partial t}+ \frac{\partial (U_j U_i)}{\partial x_j} \;=\;-\;\frac{\partial
  P}{\partial x_i}\;+\;\;\frac{1}{Re_{\tau}} \frac{\partial^2
  U_i}{\partial x_j \partial x_j} -\frac{\partial \langle u_j u_i
  \rangle }{\partial x_j},
\end{equation}

\noindent where $U_i$ (equivalent to U, V, and W), is the mean velocity and $P$ the mean pressure, $x$ and $t$ are the space and time coordinates, respectively, and $R_{\tau}$ is the friction Reynolds number $u_{\tau}
\delta/ \nu$. All variables are non-dimensionalized using the friction velocity, $u_{\tau}$ and half the distance between walls, $\delta$, which represent the characteristic velocity and length scales. Furthermore, when the original Navier-Stokes equations
are time-averaged, a tensor arises consisting of new unknowns, known
as the Reynolds stress tensor, $ \langle u_j u_i \rangle $. Given that
this tensor is symmetric, for solving these equations there are 6 new 
unknowns, in addition to $U_i$ and P.

Previous task involves incorporating a procedure or method for
computing these 6 new unknowns (the velocities U, V, and W and
pressure P are found using the conservation equations of momentum and
mass). These techniques are called RANS modeling for turbulence, as
commented above, with the Kappa-Epsilon model being an example of
them. In general, these models employ a pseudo viscosity that is
determined based on the influence of turbulence. This approach draws
an analogy between the behavior of a Newtonian fluid and that of a
turbulent flow. By incorporating this pseudo viscosity, the models aim
to capture the characteristics and dynamics of turbulent flows in a
manner similar to how a Newtonian fluid behaves.

RANS models depend on empirically-determined constants that are
calibrated using experimental turbulent flows or more accurate
techniques such as Large Eddy Simulation (LES) or Direct Numerical
Simulation (DNS). However, when these models are utilized beyond the
range of their calibrated constants, their predictions may lack
accuracy. This represents the primary challenge associated with RANS
techniques.

Following a prolonged period of stagnation, RANS modeling is now
entering a new phase with the integration of Artificial Intelligence
(AI) techniques. AI involves a diverse range of data-driven
algorithms, ranging from well-known methods like linear regression to
more advanced concepts like neural networks. With the availability of
high-fidelity data, AI allows for the capturing of potentially complex
relationships between turbulence mean-flow characteristics (features)
and modeling terms (predictions)
 
In the last years AI has been widely used in RANS models, employing a
variety of methods for addressing a wide range of problems
(\cite{milano2002, lecun2015, ling2015, ling2016a}). One such method
that has gained attention for its flexibility and precision is deep
learning (DL), which involves transforming input features through
multiple layers of nonlinear interaction (\cite{ling2016b}).

Turbulence, however, is a highly complex phenomenon that often proves
challenging to accurately model using conventional AI techniques like
deep learning (DL). One of the primary reasons for this complexity
lies in the non-local effects of turbulence. The Reynolds stresses or
turbulence stresses not only depend on the flow characteristics at the
specific location where they are calculated but also on the flow
characteristics in the surrounding regions. This interdependency poses
a significant bottleneck in turbulence modeling. To address these
non-linear problems more effectively, recurrent neural networks (RNN)
are better suited due to their ability to capture spatial or temporal
dependencies.

In particular, Long Short-Term Memory (LSTM) recurrent neural network
is currently regarded as one of the most interesting types of neural
networks, with potential for effectively capturing the non-locality of
turbulence. LSTMs have found widespread
applications in various fields, including language modeling, sentiment
analysis, machine translation, speech recognition, and time series
forecasting. In engineering, LSTMs are particularly useful for
predicting trends and patterns in time series data, such as stock
prices, weather patterns, electricity demand, anomaly detection in
sensor data, industrial machinery health monitoring, and equipment
failure prediction, among others  (\cite{schmidhuber, sutskever}). LSTMs excel in tasks where data has
a sequential nature and long-term dependencies need to be captured,
making them a promising approach for turbulence modeling.

\begin{figure}[ht] 
{\par\centering
  \resizebox*{0.7\columnwidth}{!}{\includegraphics{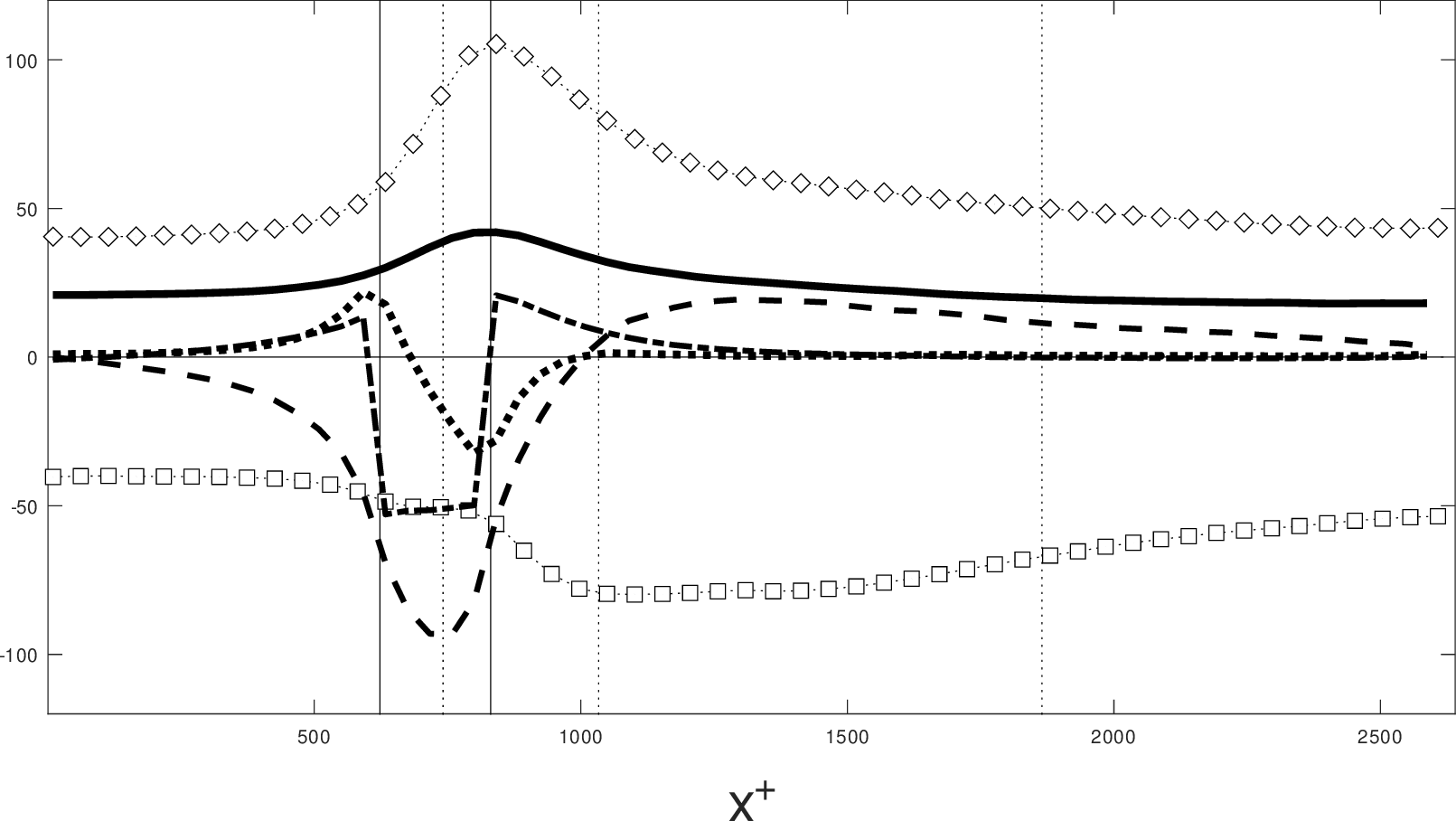}}
  \par}
\caption{\small{Mean flow and Reynolds stress dimensionless
    longitudinal profiles from DNS data for a channel flow with
    $Re_{\tau}=300$, perturbed with an adverse pressure gradient at a
    narrow region of wide $W^+=220$ at the buffer region. Vertical
    solid lines are the perturbation region limits, and the 3 vertical
    broken-lines, show middle, $W^+$ and $5W^+$ distance from
    perturbation region.  Solid line, $\partial U^+/\partial y^+$; \lines
    $\partial P^+/\partial y^+$; \brokenline, $20 \times \partial
    U^+/\partial x^+$; \bullets, $20 \times \partial V^+/\partial x^+$; \boxs
    \normalsize, $50 \times \langle uv \rangle^+$; \diamonds $10 \times
    \langle uu \rangle^+$ .}}
\label{LSTMforRANS1} 
\end{figure}

In this study, an application of LSTM is performed to predict Reynolds
stresses from data generated by Direct Numerical Simulation (DNS). The
following section discusses aspects related to the Reynolds stress
prediction using RNN; section 3 presents and discusses some of the
LSTM models tested in the study; section 4 gives some numerical
details of the RANS simulations; section 5 discusses and shows the
results of the neural network fitting and propagation of predictions
through RANS simulations, and finally section 6 lists some
conclusions.

\section{Reynolds stress prediction with LSTM RNN}

\begin{table}[ht]
\begin{center}
\begin{tabular}{lllll}
\hline \hline

Name       &$R_{\tau}$& Perturbation      &Parameter                        & Purpose     \\
\hline
Developed  & 300      &                   &                                 & train. test.\\
BRE3006    & 300      & blowing           &$v^+=0.6$                        & train. test.\\
SRE3006    & 300      & suction           &$v^+=0.6$                        & train. test.\\
APGRE30025 & 300      & adv. p. grad. step&$\partial P^+/\partial x^+=+0.25$& train. test.\\
FPGRE30025 & 300      & fav. p. grad. step&$\partial P^+/\partial x^+=-0.25$& train. test.\\
BRE3004    & 300      & blowing           &$v^+=0.4$                        & val.        \\
SRE3004    & 300      & suction           &$v^+=0.4$                        & val.        \\
APG100045  & 600      & adv. p. grad. step&$\partial P^+/\partial x^+=+0.40$& val.        \\
FPG100045  & 600      & fab. p. grad. step&$\partial P^+/\partial x^+=-0.40$& val.        \\
Developed  & 1000     &                   &                                 & val.        \\
\hline
\end{tabular}
\end{center}
\caption[table1]{\label{table1}\small Data set for training, testing  or validation. $v^+$ is the dimensionless wall-normal velocity in the slot at the wall; $\partial P^+/\partial x^+$ is the pressure gradient step in the perturbation region.} 
\label{tabla} 
\end{table}

One of the most challenging aspects of turbulence is its non-local
nature. Let's consider a scenario where we have a statistically
stationary turbulent flow (the mean flow characteristics remain
constant with time). When this flow is disturbed by an immersed body,
for example, the primary effect is on the pressure field, mainly
upstream and around the body. These pressure field changes
subsequently influence the mean velocity field around the object,
resulting in an increase in turbulence primarily downstream of the
body. In other words, the modifications in mean pressure and velocity,
and turbulence fields, occur in different regions of the flow,
emphasizing the non-local characteristics of turbulent flows.

Figure \ref{LSTMforRANS1} shows the longitudinal profiles of mean flow
and Reynolds stress in a channel flow perturbed by an adverse pressure
gradient in a narrow region near the wall. This figure reveals that
the maximum values of the shear Reynolds stress $\langle uv \rangle$
occurs downstream of the perturbation region, which is clearly
downstream where the maximum changes in the mean pressure and velocity
are observed.

In the context of RANS turbulence models, such as the
$\kappa-\epsilon$ model (\cite{pope2000}), the inclusion of non-local effects is
achieved through the convection and diffusion of kinetic energy
($\kappa$) and turbulence dissipation ($\epsilon$). However,
incorporating these non-local effects becomes more challenging when
using a regression model in deep learning (DL). In this case, it is
more appropriate to utilize a recurrent neural network such as LSTM.

Another aspect of an NN model for Reynolds stress is its
universality. The question arises regarding the level of universality
and complexity the model should possess, or conversely, how
specialized and simple it can be. Developing a universal model with an
NN would necessitate a large number of adjustable parameters to
capture all possible flow characteristics. Furthermore, fine-tuning an
NN with a high parameter count is exceedingly challenging, and
integrating a neural network with a large number of parameters into a
computational fluid dynamics (CFD) software presents a significant
obstacle due to the substantial computational demands it entails.

Consequently, this study aims to utilize an RNN with a reduced
parameter count for rapid prediction in a RANS simulation. To achieve
this, the LSTM neural network was trained, tested, and validated, using data generated from
Direct Numerical Simulations (DNS) of fully developed or perturbed (developing) turbulent flows. All the flows
considered in this study are statistically 2D channel flows,
comprising the following scenarios: a) developed flows; b) flows
perturbed with wall injection; c) flows perturbed with wall suction;
d) flows perturbed with an adverse pressure gradient in a localized
region; and e) flows perturbed with a favorable pressure gradient in
a localized region. Each perturbed flow initially represents a
developed flow, with a small region of width W+=220 being
perturbed. To generate boundary conditions for a perturbed flow,
two parallel DNS simulations were conducted, with the first simulation
employing periodic boundary conditions. For more detailed information,
please refer to the bibliography for additional insights
(\cite{pasinato2012}). Table \ref{tabla} provides a comprehensive list of the flows utilized in the study, accompanied by pertinent details such as perturbation parameters and their respective purposes.

Given that the turbulent flows studied are statistically stationary,
the neural network is trained using space-sequences instead of
time-sequences. Each space-sequence consists of 64 consecutive values
along the longitudinal direction, with an approximate dimensionless
distance of $X^+ \simeq 41$ between each value. Consequently, the
complete sequence covers a distance of approximately $X^+ \simeq
2620$, representing the entire physical domain. In other words, if for a DNS with $R_{\tau}=300$ a numerical grid with $N=254$ nods is used in the longitudinal direction, for each spatial sequence N/4 nodes of the DNS solution are employed at a fixed $y^+$ distance from the wall.

To ensure a diverse representation during training, the full sequences
are randomly shuffled. Additionally, the values are normalized between
0 and 1, using the global maximum and minimum values of the data set. This
normalization process enables consistent scaling across different
variables. Finally, the entire data set is exclusively divided into
training and testing data, with the testing data comprising $20\%$ of
the complete data set.

\section{LSTM models}

\begin{table}[ht]
\begin{center}
\begin{tabular}{lllll}
\hline \hline

Name          &  Parameters           &  Learning rate & Features   & Predictions\\
\hline
L110DL           & 531          & 0.025-0.001  & $S_{11}^+$; $S_{12}^+$     & $\langle uv \rangle^+ $\\
L15L25DL         & 386          & 0.025-0.001  & $S_{11}^+$; $S_{12}^+$     & $\langle uv \rangle^+ $\\
L15DL-Y          & 186          & 0.025-0.001  & $S_{11}^+$; $S_{12}^+; Y$  & $\langle uv \rangle^+ $\\
L110DL-Y         & 571          & 0.025-0.001  & $S_{11}^+$; $S_{12}^+; Y$  & $\langle uv \rangle^+ $\\

\hline
\end{tabular}
\end{center}
\caption[table]{\label{table1}\small LSTM tested models. }
\label{table2} 
\end{table}

\begin{figure}[ht] 
{\par\centering
  \resizebox*{0.6\columnwidth}{!}{\includegraphics{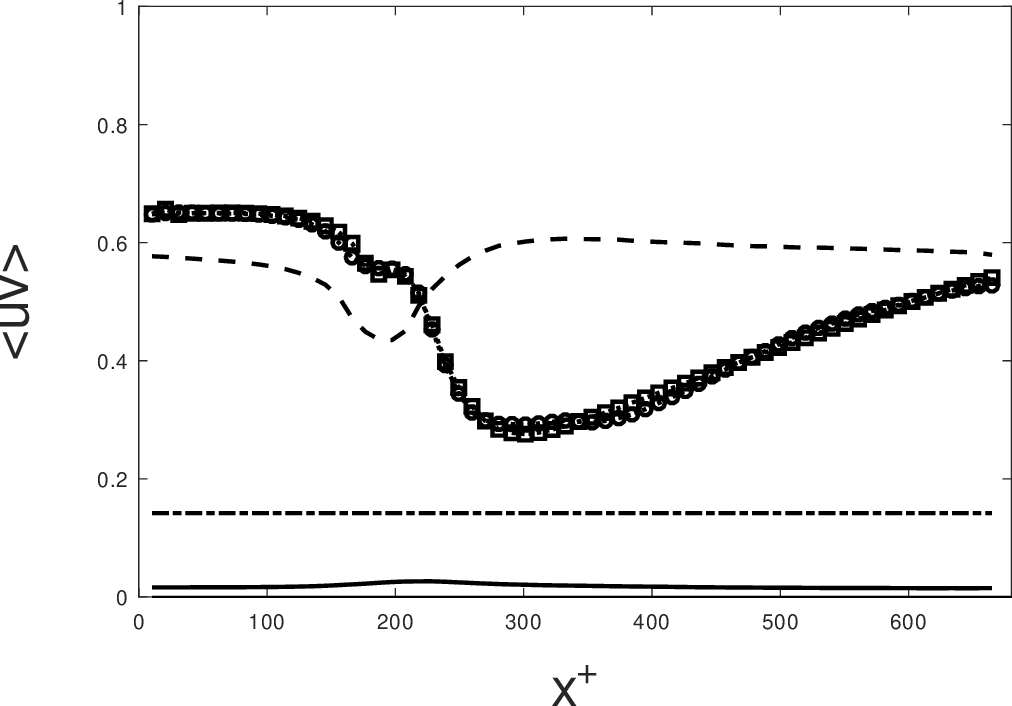}}
  \par}
\caption{\small{Prediction of $\langle uv \rangle^+$ with the LSTM model
    L110DL-Y.  Solid line, $S_{12}^+$; \brokenline, $S_{11}^+$; \lines,
    $y/\delta$; \boxs, \normalsize Prediction; \circs, DNS data. The data set
    was normalized in the 0-1 range with the global maximum and minimum.}}
\label{LSTMforRANS10} 
\end{figure}

\begin{figure}[ht] 
{\par\centering
  \resizebox*{0.65\columnwidth}{!}{\includegraphics{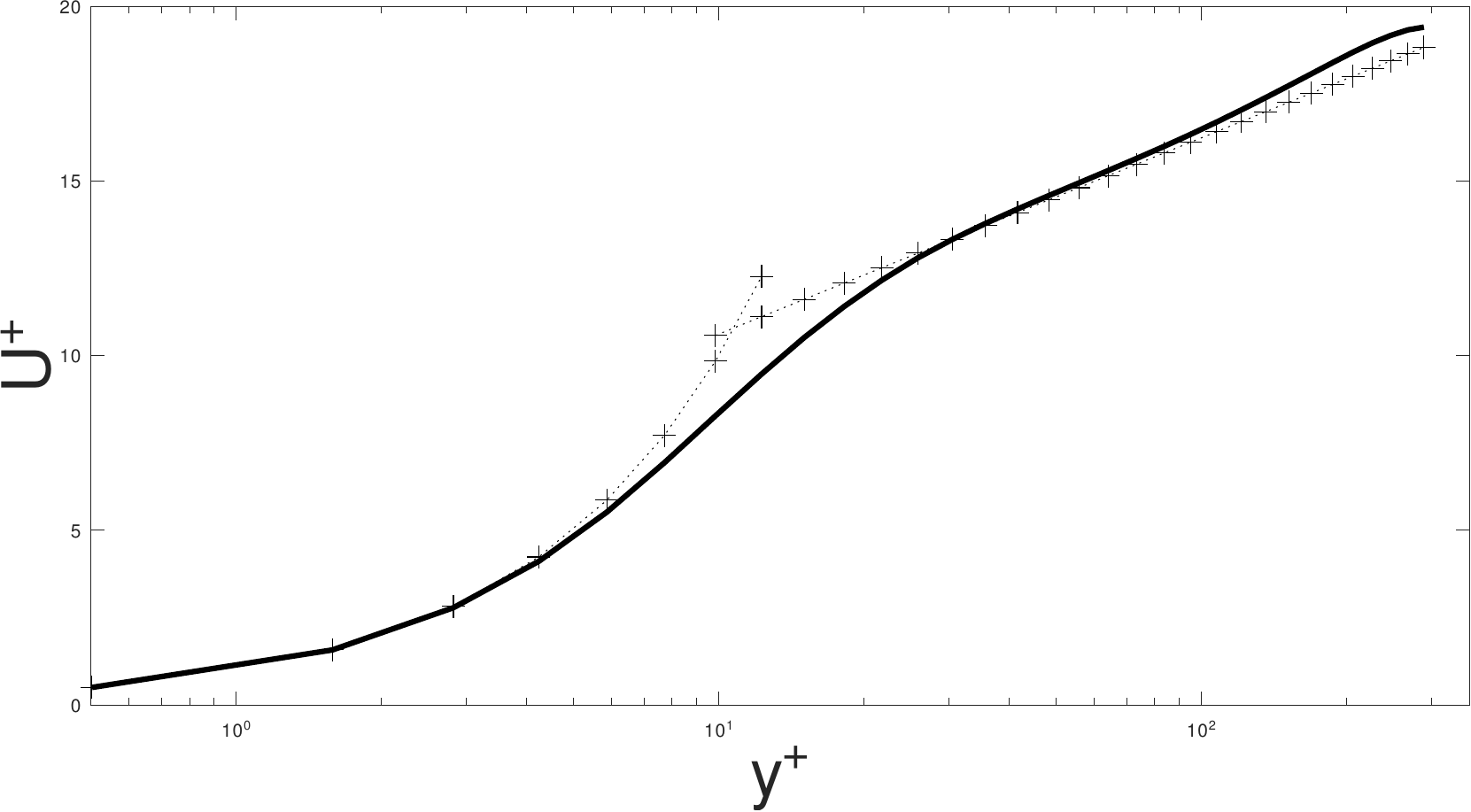}}
  \par}
\caption{\small{Mean velocity for a developed channel flow with
    $Re_{\tau}=300$ with $\langle uv \rangle^+$ predicted from the LSTM
    model L110DL-Y. Solid line, $U^+$; \pluss, Logarithmic profile
    $1/0.41\; ln(y^+)+5.5$.}}
\label{LSTMforRANS3} 
\end{figure}
 
\begin{figure}[ht] 
{\par\centering
 \resizebox*{0.8\columnwidth}{!}{\includegraphics{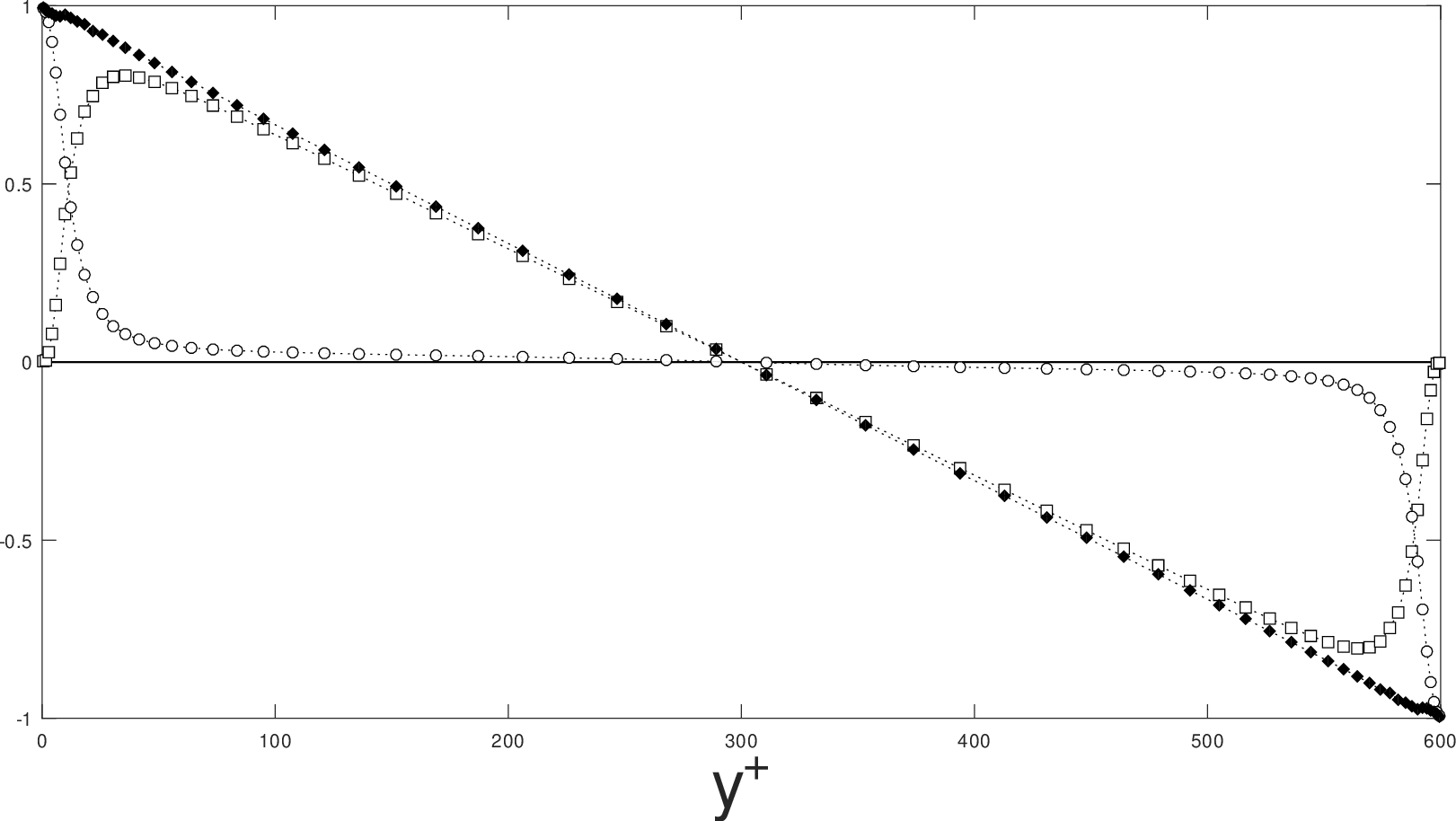}}
  \par}
\caption{\small{Shear stresses for a developed channel flow with
    $Re_{\tau}=300$ with $\langle uv \rangle^+$ predicted with the LSTM
    model L110DL-Y. Filled diamonds, total stresses; \boxs,
    \normalsize $\langle uv \rangle^+$; \circs, molecular stress $dU^+/dy^+$.}}
\label{LSTMforRANS5} 
\end{figure}

All LSTM models were implemented in two different numerical codes: one
developed in-house using modern Fortran, and the other implemented in
Python using the Keras-TensorFlow libraries (\cite{tensorflow2015-whitepaper}). The porpouse of the
Fortran code was to gain insight into the forward and backward
algorithms of LSTM and compare them with the Keras version (
\cite{chen2018}).

The LSTM, like other deep learning models, has three 
hyper-parameters: a) the number of stacked layers (or stacked LSTM cells), b) the number
of memory units in each LSTM layer, and c) the learning rate. In this
study, these hyper-parameters were selected through trial and
error. 

It is worth mentioning that the TensorFlow libraries offer a wide
range of optimization methods, including dynamic learning-rate
techniques that speed up convergence. However, both the Fortran and
Keras codes faced challenges when optimizing models with many stacked
layers (5 or more layers) and numerous memory units. To simplify the
models and reduce the number of parameters, a maximum of 3 stacked
LSTM layers with a maximum of 10 memory units were used for all tested
models. Finally the best predicted solution were from models with 1
LSTM cell and 10 memory units for the data set used.

Table \ref{table2} presents some of the LSTM models tested in this study. The name $L15L25DL$ indicates an LSTM with 2 stacked layers, each having 5 memory units, and a dense output layer. The components $S_{11}^+$ and $S_{12}^+$ represent the dimensionless mean rate-of-strain-tensor $S_{ij}$, which are calculated as $S_{ij}^+=1/2(\partial U^+_i/\partial x^+_j+\partial U^+_j/\partial x^+_i)$. Additionally, $Y$ denotes the dimensionless wall distance ($y/\delta$). It is important to note that the axes in $S{ij}$ are non-dimensionalized differently from the distance to the wall $Y$.

Although the primary objective of this study was to predict the shear
Reynolds stress $\langle uv \rangle$, some of the models were
developed to predict the normal Reynolds stress $\langle uu \rangle$,
as an initial step, followed by an LSTM model to predict $\langle uv
\rangle$ based on the predicted $\langle uu \rangle$. In all of these
cases, the optimization of the LSTM models for $\langle uu \rangle$
yielded superior and faster results compared to the LSTM models for
$\langle uv \rangle$.

The previous architecture of one RNN to predict $\langle uu \rangle$ and a second RNN to predict $\langle uv \rangle$ is based on the non-linearity of
turbulence. In the context of channel turbulent flows, it is
well-established that turbulence acquires energy from the mean flow
first through $\langle uu \rangle$. Subsequently, the instantaneous
pressure field transfers a portion of this energy to the $\langle vv
\rangle$ and $\langle ww \rangle$ components (\cite{tennekes}). As
consequence as depicted in Fig. \ref{LSTMforRANS1}, the modification
of $\langle uu \rangle$ occurs promptly after the mean velocity and
pressure field are perturbed, while the alteration of $\langle uv
\rangle$ experiences a certain delay. No results are presented here of
the LSTM model of $\langle uv \rangle$ based on a previous prediction of
$\langle uu \rangle$.
  
\section{Numerical details of the RANS simulations}

The RANS simulations with the propagation of the LSTM predictions and
the $\kappa-\epsilon$ model were computed with a physical domain of
$3\pi \delta \times 2\delta \times 4/3\pi \delta$, and a numerical
grid of $64 \times 64 \times 64$ ($\Delta x^+=44$, $\Delta z^+=19.6$,
$\Delta y^+_{max}=26.4$, and $\Delta y^+_{min}=1$.), and the time step
$0.15 \nu/u_{\tau}^2 $. In the wall-normal direction a non-uniform
mesh distributed with the hyperbolic tangent function is used and the
expansion ratio is adjusted to ensure that the $y^+$ of the first cell
center is equal to 1. Furthermore, a van Driest function near the wall
is used with the $\kappa-\epsilon$ model. Considering the low Reynolds
number flows employed in this study, it is believed that the wall
modeling approach adopted is adequate.

The RANS simulation of perturbed flow follows a similar approach to DNS. Initially, a fully developed flow is considered, and a small region of width W+=220 is perturbed. To generate boundary conditions for the perturbed flow, a second simulation is conducted in parallel with periodic boundary conditions, representing the fully developed flow. In other words, at every time-step, the input boundary condition for the perturbed flow is extracted from the middle of the computational domain of the fully developed flow.

\section{Results and Discussion}

\begin{figure}[ht] 
{\par\centering
  \resizebox*{0.65\columnwidth}{!}{\includegraphics{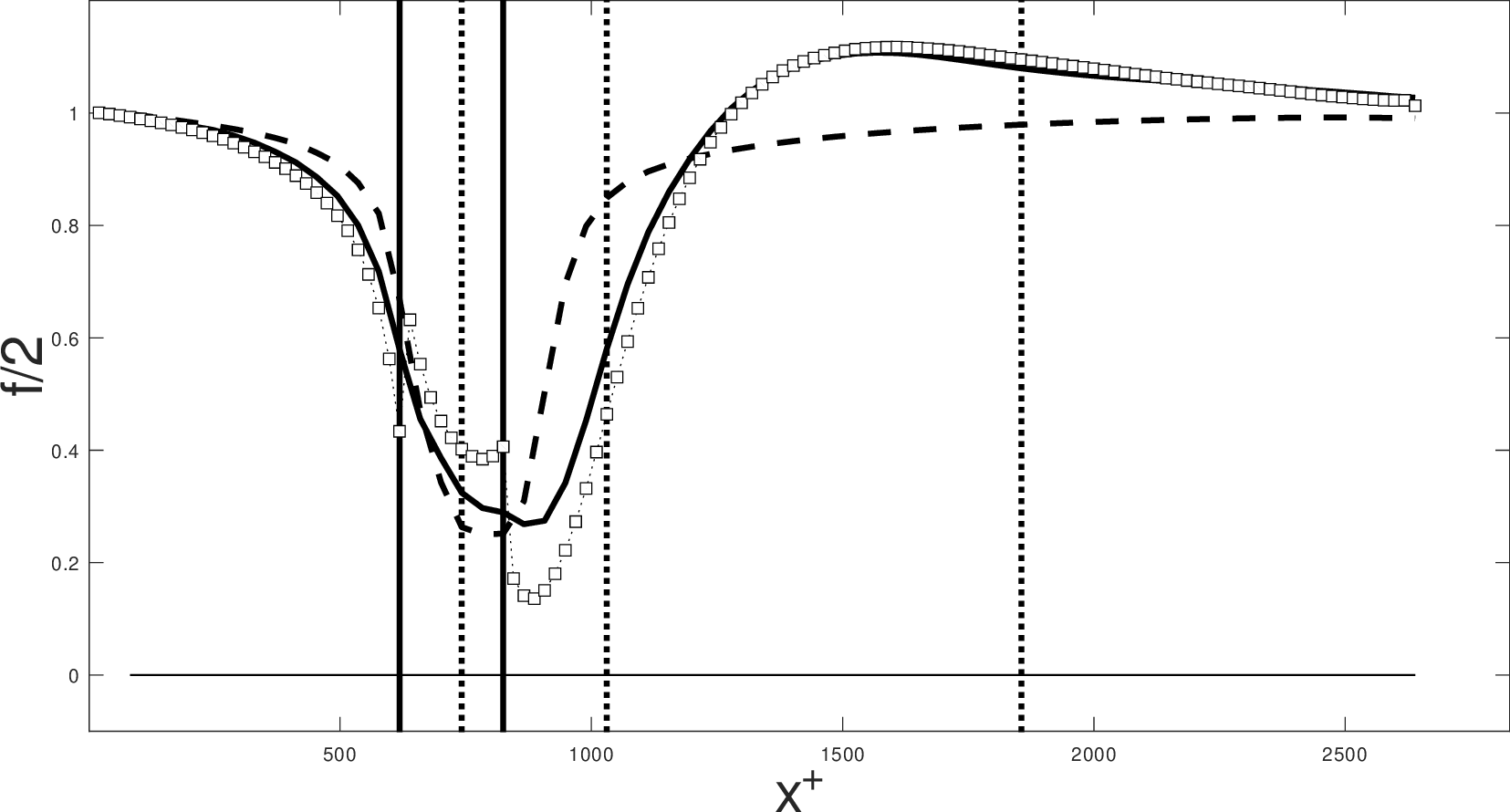}}
  \par}
\caption{\small{Comparison of half of the skin-friction for a channel flow
    with $Re_{\tau}=300$ perturbed with blowing with an injection
    dimensionless velocity $0.40$ from a narrow slot at the wall, with
    $\langle uv \rangle^+$ predicted with the LSTM L110DL-Y model. Solid
    line, LSTM; \brokenline, \normalsize $\kappa-\epsilon$ model;
    \boxs, \normalsize DNS.}}
\label{LSTMforRANS6} 
\end{figure}
 
In this section, we present the results of the LSTM predictions and
their integration with a CFD code for propagation. The term
"propagation" refers to the utilization of the LSTM RNN as a
substitute for a RANS turbulence model. At each time-step, sequences
are formed and the LSTM model is employed to calculate the Reynolds
stress $\langle uv \rangle$ at every node of the grid. While we
provide some comparisons between the LSTM prediction and results
obtained from the $\kappa-\epsilon$ model (which computes all shear and
normal Reynolds stresses), the RANS simulations with LSTM predictions
only consider the shear Reynolds stress $\langle uv \rangle$. The
remaining Reynolds stresses were taken equal to zero.

As mentioned earlier, the purpose of using the LSTM RNN in this study
is not to capture the universal behavior of turbulence, but rather to
predict statistical stationary 2D turbulent flows specifically for
channel flow. The LSTM model serves as a proof of concept for
predicting Reynolds stresses. Consequently, the turbulent flows
utilized for validating the LSTM model share certain similarities with
the flows employed for model adjustment and testing.

For instance, all the turbulent flows used for model adjustment,
testing, and validation pertain to channel flows. These flows
encompass developed or perturbed developing flows, characterized by
low or medium friction Reynolds numbers $R_{\tau}$ (300-1000). It is
worth noting that the mean velocity and normal Reynolds stresses,
along with other statistical measures, exhibit low Reynolds number
effects in these flows. The developing flows represent turbulent flows
that have been perturbed from the wall or in small regions in the
buffer region.

\begin{figure}[ht] 
{\par\centering
  \resizebox*{0.65\columnwidth}{!}{\includegraphics{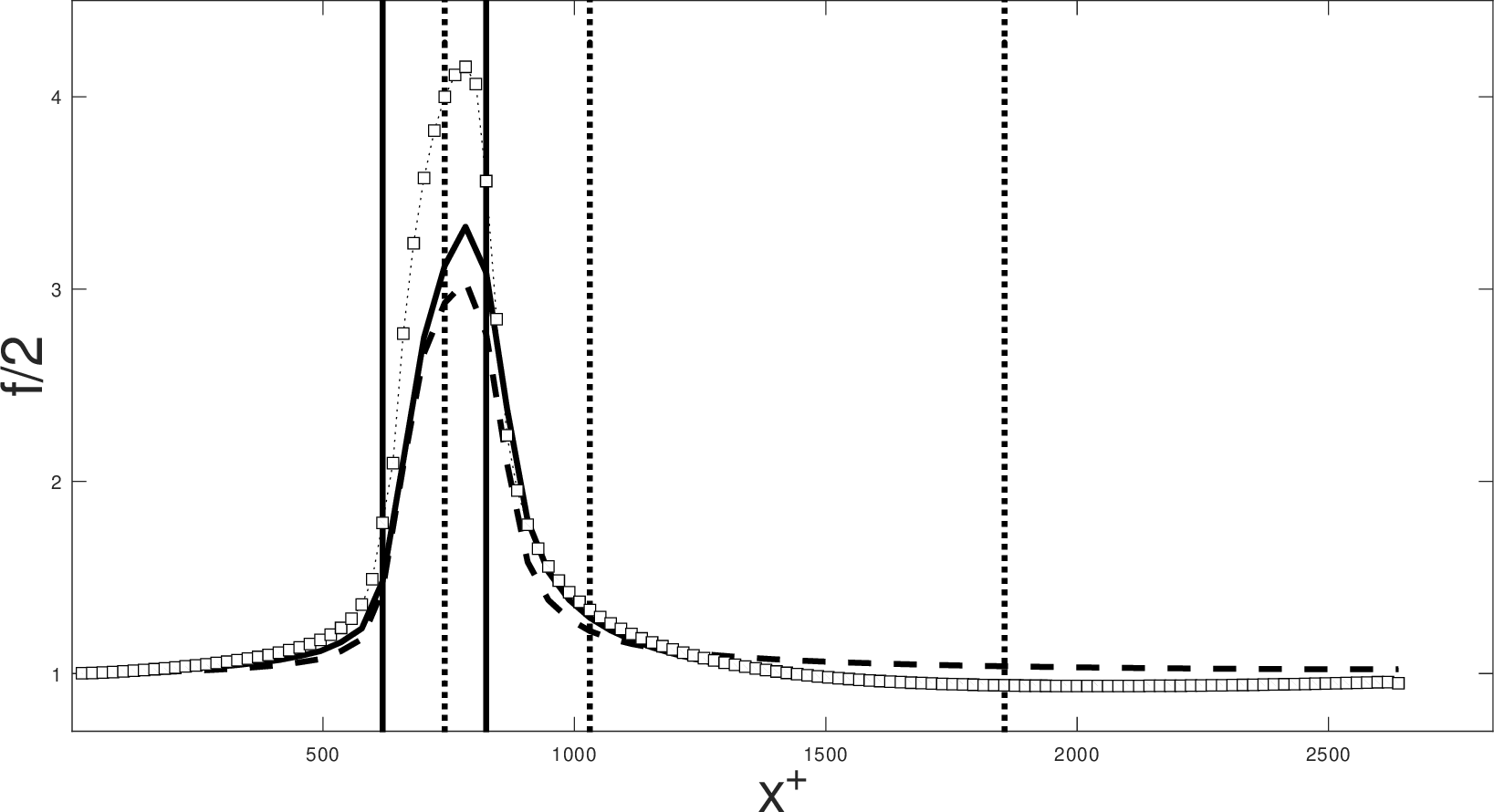}}
  \par}
\caption{\small{Comparison of half of the skin-friction for a channel flow
    with $Re_{\tau}=300$ perturbed with suction from a narrow slot at
    the wall, with $\langle uv \rangle^+$ predicted with the LSTM
    L110DL-Y model. Solid line, LSTM; \brokenline, \normalsize
    $\kappa-\epsilon$ model; \boxs, \normalsize DNS.}}
\label{LSTMforRANS7} 
\end{figure}

\begin{figure}[ht] 
{\par\centering
  \resizebox*{0.65\columnwidth}{!}{\includegraphics{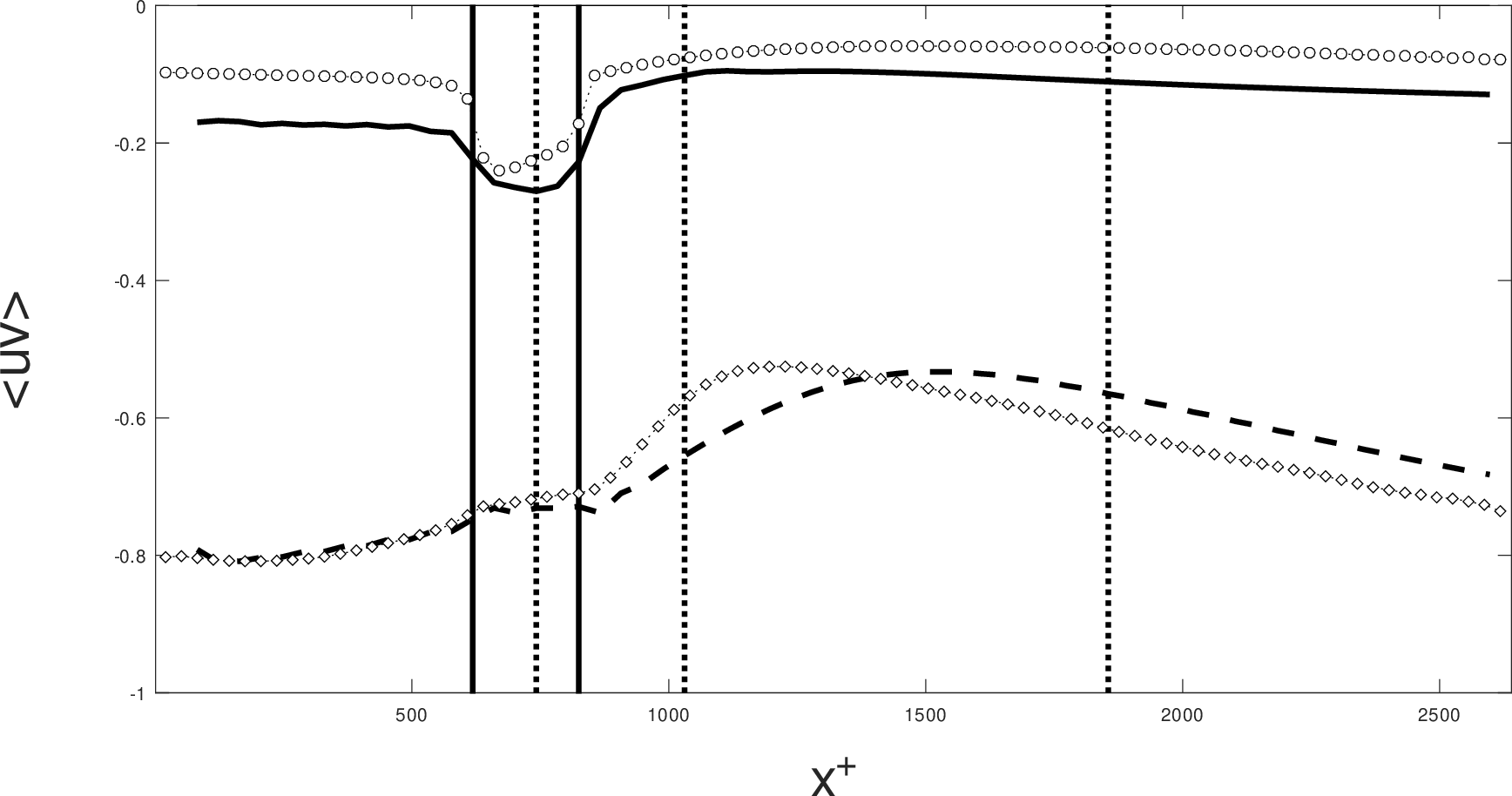}}
  \par}
\caption{\small{Comparison of the shear Reynolds stress for a channel
    flow with $Re_{\tau}=300$ perturbed with suction from a narrow
    slot at the wall, with $\langle uv \rangle^+$ predicted from the
    L110DL-Y model. $y^+=5$, solid line, LSTM; \circs DNS; $y^+=30$,
    \brokenline, LSTM; \diamonds, DNS.}}
\label{LSTMforRANS9} 
\end{figure}

\vskip0.2cm
\noindent{\it Prediction of the shear Reynolds stress without
  propagation:} Figure \ref{LSTMforRANS10} illustrates the LSTM-based
prediction of the shear Reynolds stress $\langle uv \rangle^+$ utilizing
$S_{11}^+$, $S_{12}^+$, and Y($=y/\delta$) as input parameters, with L110DL-Y
model. The LSTM model employed for this prediction follows a
sequence-by-sequence (seqxseq) architecture (\cite{sutskever}). To ensure consistent
scaling, all data points have been normalized within the range of 0-1,
utilizing the global minimum and maximum values across the entire
dataset. These normalized values are subsequently transformed back to
their original, real-world scale using the same minimum and maximum
values.

\vskip0.2cm
\noindent{\it Propagation of predicted shear Reynolds stress in
  developed flow:} In fully developed turbulent flows, the momentum equation (\ref{eq102}) in the longitudinal direction can be expressed in dimensionless form using the characteristic velocity scale $u_{\tau}$ and boundary layer thickness $\delta$, as mentioned earlier. It takes the following form: 
	
\begin{equation}\label{eq103}
\frac{d U^+}{dy^+}\;=\; \langle uv \rangle^+ + (1-\frac{y^+}{R_{\tau}})
\end{equation}

\noindent where the pressure gradient has been substituted with the wall shear stress (\cite{tennekes}). Equation (\ref{eq103}) illustrates the equilibrium between the Reynolds shear stress and the wall-normal gradient of the longitudinal mean velocity for a fully developed turbulent flow. This equilibrium is depicted in the plots of $U^+$, $\langle uv \rangle^+$, and $dU^+/dy^+$ as functions of the dimensionless distance to the wall, $y^+$, in Figs. \ref{LSTMforRANS3} and \ref{LSTMforRANS5}. Specifically, the agreement of the $U^+$ profile with the logarithmic law of the wall in the logarithmic sub-region, shown in Fig. \ref{LSTMforRANS3}, indicates that the predicted value of $\langle uv \rangle^+$ is accurate.

Thus, as observed in these figures, the mean velocity exhibits a logarithmic
sub-region with appropriate values for the given Reynolds number,
indicating the characteristic velocity profile of a developed
turbulent flow. Furthermore, the shear stresses, including both
molecular and turbulent components, demonstrate a state of perfect
equilibrium under developed flow conditions. This implies that the
LSTM predictions successfully capture the essential features of
developed turbulent flows.

In order to evaluate the stability of the results
obtained from the LSTM model, a series of tests were conducted using
flows with initial conditions significantly distant from a fully
developed state. The objective of these tests was to assess the
capability of the LSTM model to converge towards a stable solution.

\vskip0.2cm
\noindent{\it Propagation of predicted shear Reynolds stress in
  developing flows:} In this section the results of the LSTM model
L110DL-Y propagated in developing turbulent channel flow are
presented. The first case is a blowing case in a channel with a
dimensionless injection velocity of $0.40$ (the set of data for
training included an injection case with a dimensionless injection
velocity of $0.60$). The results
of the propagated LSTM predictions are shown in comparison with
$\kappa-\epsilon$ model and data from DNS, Figs. \ref{LSTMforRANS6} to
\ref{LSTMforRANS9}.

Figures \ref{LSTMforRANS6} and \ref{LSTMforRANS7} clearly demonstrate the LSTM model's superior performance in computing the skin friction when compared to the traditional $\kappa-\epsilon$ model. Additionally, Figure \ref{LSTMforRANS9} illustrates that the propagated shear Reynolds stress follows the general trend observed in the DNS data.

A RANS prediction depends on various factors, such as the numerical grid, software precision, and, of course, the RANS model itself. In this study, all these aspects were reasonably implemented. Based on the presented comparisons between the LSTM RNN and the $\kappa-\epsilon$ model predictions, it can be concluded that LSTM-RANS exhibits superior performance. However, it is essential to note that the LSTM-RANS simulations with the L110DL-Y model used here required approximately 50\% more time for computation than $\kappa-\epsilon$ simulations.

Regarding the time-processing aspect of LSTM-RANS, it is worth mentioning that a more comprehensive study should be conducted. This broader investigation should encompass the use of different RANS models and LSTM networks with varying numbers of parameters to draw more definitive conclusions.

In summary, the study demonstrates that LSTM-RANS outperforms the traditional $\kappa-\epsilon$ model, but further research is needed to explore its efficiency and effectiveness fully.

\section{Conclusions}

In this study, we introduce the application of an artificial Recurrent Neural Network (RNN) known as LSTM as a promising alternative to a turbulent RANS model. The primary objective was to utilize the LSTM model for predicting the shear Reynolds stress in both developed and developing turbulent channel flows, and subsequently propagate this prediction.

To assess its performance, we conduct a comparative analysis, where we compare the LSTM results, propagated through CFD simulations, with the outcomes from the traditional $\kappa-\epsilon$ model and data obtained from DNS. Remarkably, the LSTM approach demonstrates a strong performance in these analyses, showing its potential as an effective technique for modeling the shear Reynolds stress in turbulent flows.

It is essential to note that this study serves as a proof of concept, and the validation process utilizes 2D statistical turbulent flows as the training and test data sets. Despite this limitation, the results presented here indicate the LSTM RNN's efficacy in modeling the shear Reynolds stresses accurately in turbulent flows.

Overall, the findings from this study open up exciting possibilities for further exploring the capabilities of LSTM models as alternatives to conventional RANS models in the context of turbulence prediction and modeling.

\bibliographystyle{unsrtnat}
\bibliography{LSTMforRANS}

\end{document}